\documentclass[english]{article}
\usepackage[T1]{fontenc}
\usepackage[latin1]{inputenc}
\usepackage{latexsym}
\usepackage{epsfig}
\usepackage{a4}
\usepackage{amsfonts}

\makeatletter

\usepackage{epsfig}
\usepackage{babel}
\makeatother
\newcommand{\beq}[1]{\begin{equation}\label{#1}}
\newcommand{\eeq}{\end{equation}}
\newcommand{\beqar}[1]{\begin{eqnarray}\label{#1}}
\newcommand{\eeqar}{\end{eqnarray}}

\newcommand{\ga}{\gamma}

\begin{document}

{\begin{center} {\LARGE { Diffractive large transferred momentum
photoproduction of vector mesons} }
\\ [8mm] {\large  A. Ivanov\footnote{e-mail:
Alexander.Ivanov@itp.uni-leipzig.de} and R. Kirschner \\ [3mm] }

 Naturwissenschaftlich-Theoretisches Zentrum und
Institut f\"{u}r Theoretische Physik, Universit\"{a}t Leipzig, \\
Augustusplatz 10, D-04109 Leipzig, Germany

\end{center}
}

\vspace{3cm} \noindent {\bf Abstract}

\noindent The large $t$ behaviour of the helicity amplitudes of diffractive
photoproduction is estimated relying on models of the photon and meson
light-cone wave functions and on the double-logarithmic approximation to the
exchanged gluon interaction. The role of large-size colour dipole
contributions to the photon-meson transition impact factor is discussed.
\newpage

\section{Introduction.}

Hard diffractive vector meson production is one of the topics considered in
the analysis of HERA experiments, \cite{Abramowicz:1998ii}. Data on
photoproduction are available
\cite{Chekanov:2002rm,Aktas:2003zi,Brown:2003bi,Breitweg:1999jy}
extending to relatively large $t.$

Diffractive photoproduction at relatively large momentum transfer
is a particular example of semi-hard processes determined by two large
scales, $s \gg -t \gg m_V^2 $.

 Because of the large momentum transfer one
expects that an essential part of the interaction can be described by
perturbative QCD, in particular the BFKL approach should be applicable
for calculating the diffractive exchange. The intriguing question is whether
the coupling of this perturbative exchange to the scattering particles, the
photon - meson impact factor, is dominated by short distance configurations.

In the case of $J/\Psi$ production at large $t$ the heavy quark mass
guarantees the applicability of the perturbatively calculated impact factor
\cite{Forshaw:1995ax,Bartels:1996fs}. In the present paper we address
the question to what extend a
perturbative calculation can represent the photon - light meson diffractive
transition, in particular, whether  this process can be treated in
the picture of small-size dipole interaction.

The case of light meson diffractive photoproduction at large $t$ has been
considered in a number of papers
\cite{Ginzburg:1996vq,Nemchik:1996pp,Ivanov:2000uq,Forshaw:2001pf,Enberg:2003jw}.
The small dipole contributions to the impact factor of all helicities
have been calculated in \cite{Ivanov:2000uq}
 by using distribution amplitudes for
both the photon and the vector meson.  It has been suggested that
the experimentally observed
dominance of the transversely polarized meson production may be explained by
a sizable chirally-odd contribution in the photon and meson light-cone
wave functions.
Calculating  the exchange by the leading $\ln s$ BFKL equation
allows to describe the $t$ dependence of the photoproduction cross sections
\cite{Forshaw:2001pf}.
The BFKL formulation of the helicity ampltudes has been presented in
\cite{Enberg:2003jw}, and its phenomenological consequences in the next article of the same authors
\cite{Poludniowski:2003yk}.

The aim of the present paper is to emphasize the role of the large dipole
size in the photon-meson transition impact factor. We adopt an ansatz for
the meson light-cone wave function used in previous studies of diffractive
electroproduction \cite{Ivanov:1998gk,Kirschner:qq,Ivanov:2003ts}
and a similar ansatz for the photon light-cone wave
function. We point out the contribution with the large momentum transfer
carried by both of the exchanged gluons, where the $q \bar q $ dipole size is
not suppressed by the large $t$.

We include the leading effect of the exchanged gluon interaction by
approximating the BFKL equation down to the lower level of double
logarithmic $\ln s \ \ln t$ accuracy \cite{Fippel:yz}.
In view of the complexity of the
amplitudes constructed from the leading $\ln s$ BFKL solution as presented
in \cite{Enberg:2003jw} our approximate treatment
is a reasonable simplification
in order to study  particular contributions. It allows to
demonstrate the main impact of the exchanged gluon interaction and to
estimate the importance of the large dipole-size contributions.

\section{Effective dipole scattering at large t.}

We recall the general approach to hard diffraction.
 The amplitude of the diffractive process
$\ga\to  V $ can be represented as the integral over the
transverse momenta of gluons in the $t-$channel (impact
representation)
$$
M^{\lambda _{i}\lambda _{f}}(s,q)=s\int ^{i\infty }_{-i\infty }
\frac{d\omega }{2\pi i} F^{^{\lambda _{i}\lambda _{f}}}(\omega,q)
\left[ \left( \frac{s}{M^{2}(m,q)}\right) ^{\omega }+ \left(
\frac{-s}{M^{2}(m,q)}\right) ^{\omega }\right], $$
\begin{equation} \label{partial_waves}
F^{^{\lambda _{i}\lambda
_{f}}}(\omega,q)= \int d^{2}\kappa d^{2}\kappa^{\prime } \Phi
^{^{\lambda _{i}\lambda _{f}}}(\kappa,q) \ {\cal G} (\kappa,
\kappa^{\prime },q,\omega ) \ \Phi ^{P}(\kappa^{\prime },q)
\end{equation} Here $q$  is the momentum transfer, $\kappa,\kappa^{\prime}$
the transversal
momenta of the exchanged gluons,  ${\cal G} $ the diffractive exchange
(Pomeron).
 $\Phi^{^{\lambda _{i}\lambda _{f}}}$ and $\Phi ^{P}$ are photon-meson
and proton
impact factors respectively.

 The photon fluctuates
into a $q\bar q$ pair long time before and this $q\bar q$ pair
converts into the vector meson long time after the interaction
with the proton.
 It is possible to represent the photon impact factor
as the convolution of the impact factor for the $q\bar q$ dipole
scattering with the light cone wave functions of the incoming
virtual photon and the  outgoing vector meson (Fig.1)
\begin{equation} \Phi^{\lambda _{i}\lambda
_{f}}(\kappa_{1},\kappa_{2})= \int
d^{2}\ell_{1}d^{2}\ell_{2}dz\Psi^{(\gamma)\lambda_{i}}(\ell_{1},z)
\phi ^{dip}(\ell_{1},\ell_{2},\kappa_{1},\kappa_{2}) \Psi
^{{V}^{\lambda _{f} *}}(\ell_{2}-zq,z), \end{equation}
\begin{eqnarray*}
   \phi^{dip}(\ell_{1},\ell_{2},\kappa_{1},\kappa_{2})=\alpha
_{s} [\delta ^{2}(\ell_{2}-\ell_{1}) +\delta
^{2}(\ell_{2}-\ell_{1}+\kappa_{1}+\kappa_{2})-  \\
   \delta
^{2}(\ell_{2}-\ell_{1}+\kappa_{1})- \delta
^{2}(\ell_{2}-\ell_{1}+\kappa_{2})] \label{dip}
\end{eqnarray*}
\begin{figure}[htbp]
\begin{center}
\vspace{-0.cm}
\epsfig{file=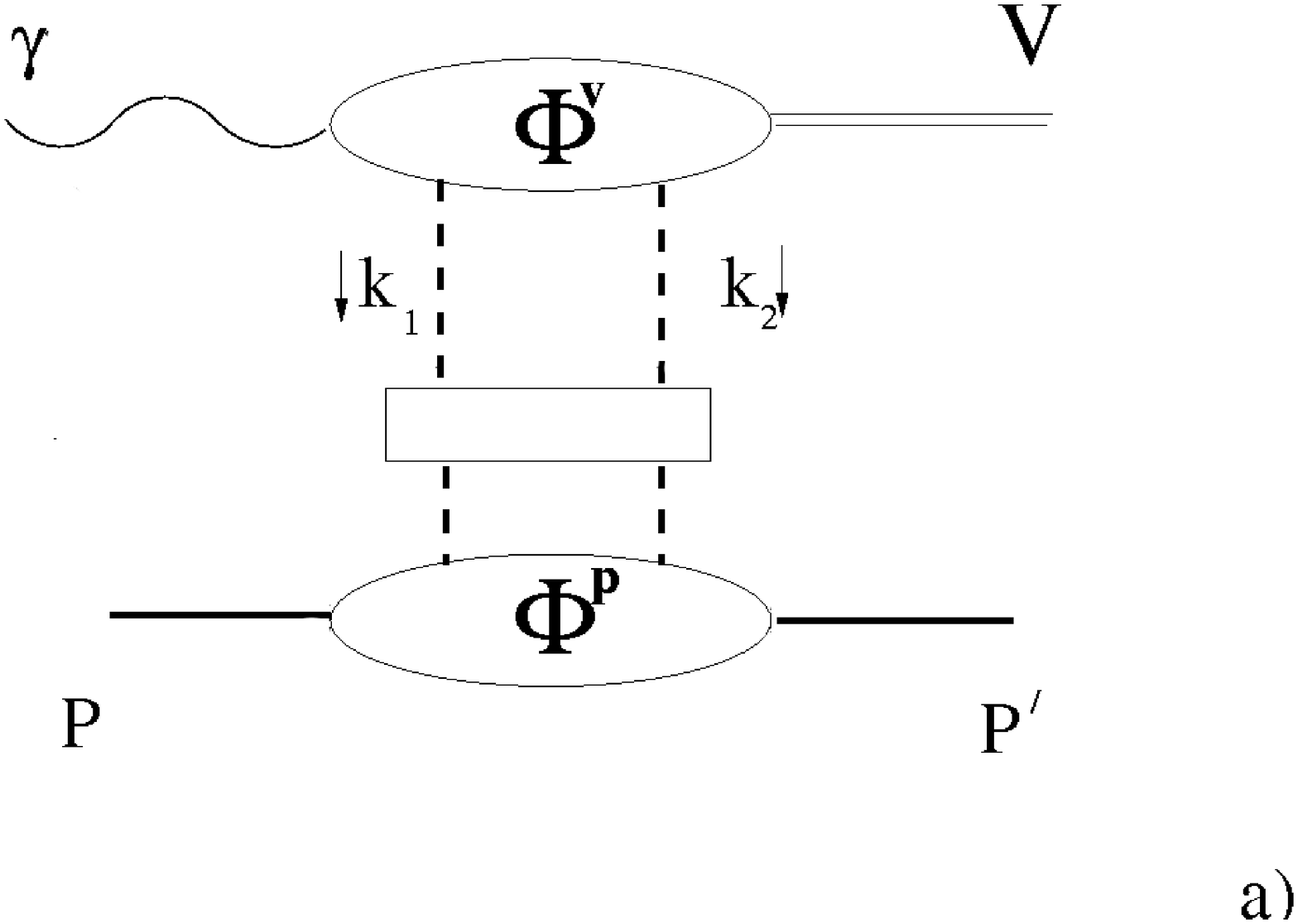,width=10cm}
\epsfig{file=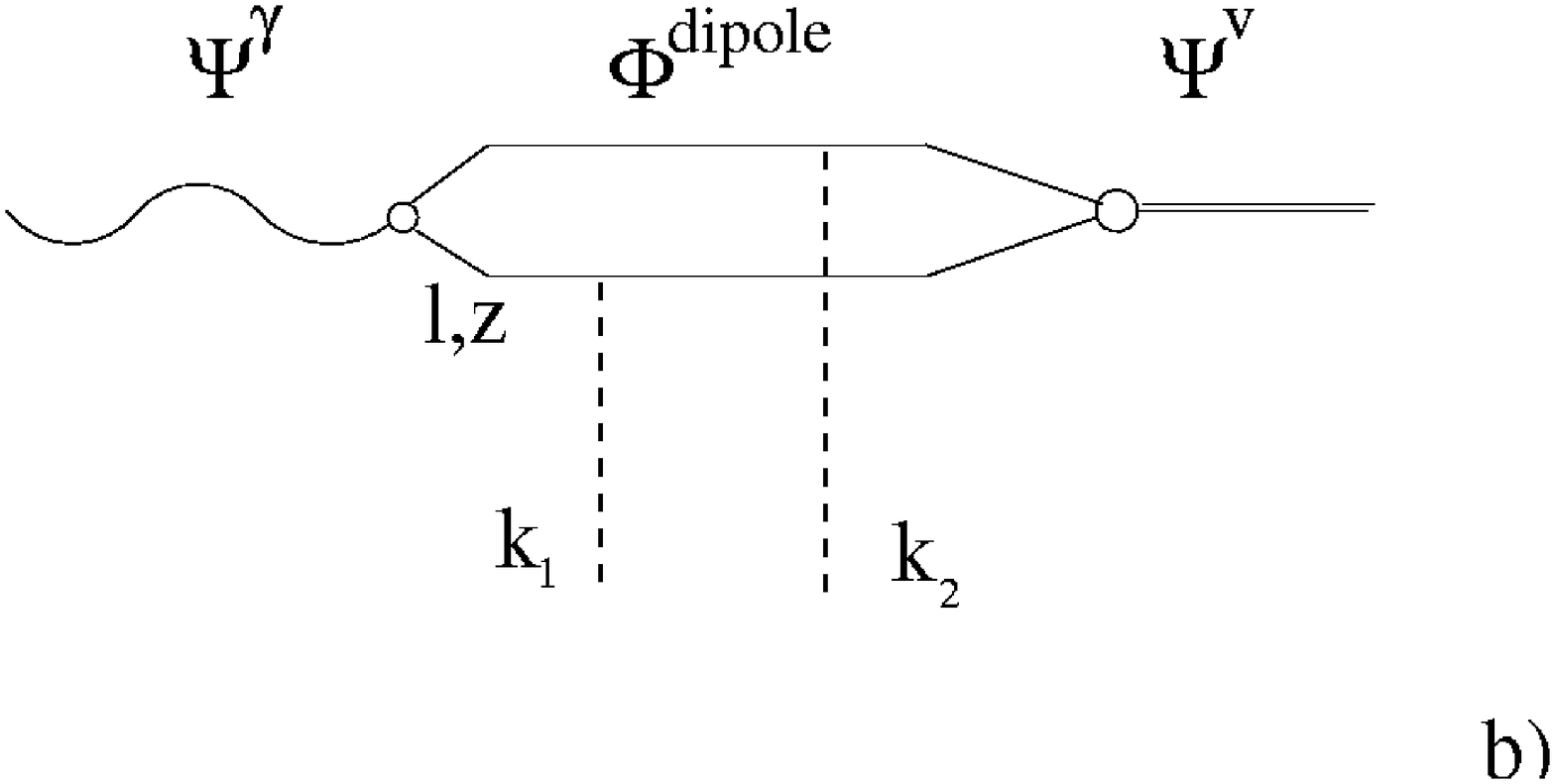,width=10cm}
\end{center}
\caption{a) Impact factor form of the $\gamma \rightarrow V $ diffractive
amplitude.
b) Contribution to the impact factor.  }
\label{fig:rt}
\end{figure}

As the photon wave function we adopt the extrapolation of the virtual photon wave function. Whereas the latter is the result of a perturbative calculation, the extrapolation to Q=0 is a model assumption
\beqar{equation}\label{psigamma}
\Psi^{(\gamma)\lambda}(\ell,z)=
\frac{V^{\lambda
}(\ell,z,0)z \bar z }{{|\ell|^{2}+m^{2}_{q}}} \cr
\ V^{(+1)}=\frac{\ell^{*}}{z},\
V^{(-1)}=\frac{\ell}{\overline{z}} \eeqar

As vector meson wave function we can use
\begin{equation}
  \Psi^{V\lambda }(\ell,z)= f_{V}\frac{V^{\lambda
}(\ell,z,m_{V})}{m^{2}_{V}} \exp\left[
-\frac{|\ell|^{2}+m^{2}_{q}}{z\overline{z}m_{V}^{2}}\right]\label{eq:psiV}\end{equation}
This form has been used earlier \cite{Ivanov:1998gk,Kirschner:qq,Ivanov:2003ts}.
It can be motivated by QCD sum rules, as resulting from the virtual photon wave function by Borel transformation and by the substitution of Borel variable by  $m_{V}^{2}$. This wave function, being close to the one of incoming photon, is a particular realization of the phenomenologically succesful concept of vector dominance.

Technically, the adopted forms of wave functions provide the advantage that the transverse momentum integration involved in the impact factor can be easily performed.

Unlike the case of electroproduction now the wave functions do not suppress the contributions from large dipole sizes. Such a suppression can result rather from the dipole impact factor involving
the large momentum transfer. Transforming the dipole impact factor to coordinate representation
we have
\[
\int
e^{i({\ell}_{1}r_{1}-{\ell}_{2}r_{2})}d{\ell}_{1}d{\ell}_{2}
\phi^{dip} ({\ell}_{1},{\ell}_{2},\kappa_1, \kappa_2)\]
\begin{equation}
=e^{izqr}\ (e^{-i\kappa_{1}r}+e^{-i\kappa_{2}r}-1-e^{i(\kappa_{1}+
\kappa_{2})r})\ \delta^{2}(r_{1}-r_{2})
\label{eq:dipole}
\end{equation}
The large momentum transfer $q$ leads to the dominance of small
dipole sizes, $r_1=r_2=O(q^{-1})$ for generic values of the
 momenta $\kappa_1, \kappa_2$ $(\kappa_1+\kappa_2=-q)$
of the exchanged gluons. This
hard contribution to the photon-meson impact factor can be
constructed with distribution amplitudes of both the photon and
the vector meson. In \cite{Ivanov:2000uq} only this contribution
has been considered. A particular feature is that one exchanged
gluon carries large and the other relatively small momentum,
$\kappa_1\ll q$ or $\kappa_2\ll q$. Eq(\ref{eq:dipole}) shows that we  have
two further regions, where the dominant dipole size is not small.
The first one is the vicinity of the end point, $z=0, z=1.$ We have
considered this contribution in the case of electroproduction
\cite{Ivanov:2003ts}. The second one corresponds to small values
of $\kappa_1+zq$ or $\kappa_1+\overline{z}q.$ This means that here the large
momentum transfer is shared by the two gluons. In this respect it is
reminiscent of the Landshoff mechanism proposed for $pp$ elastic
scattering at large $t$ \cite{Landshoff:ew}. We shall see that the
photon-meson impact factor has extra terms which contribute to
this region but are exponentially small outside of it.

\section{Impact factor $\gamma_{real}V$.} 
\setcounter{equation}{0}

First we want to consider the upper part of the
diagram. We write down the photon impact factor in the following
general form:

\begin{equation}
\Phi^{\lambda_{i}\lambda_{f}}=\int_{0}^{1}dzz \overline{z}
\varphi^{\lambda_{i}\lambda_{f}}
_{4}(z,\kappa,q)\label{eq:impact_factor}
\end{equation}

\begin{equation}
\varphi_{4}(z,\kappa,q)=\varphi(z,\kappa,q)+\varphi(z,-\kappa-q,q)-
\varphi(z,0,q)-\varphi(z,-q,q),\label{eq:4terms}
\end{equation}
where contributions of all four diagramms with different
couplings of gluons to quarks are taken into account. Summing  the diagramms with different momentum flow in effect we average over quark helecities in the quark loop.

For any contribution we have:\begin{equation}
\varphi(z,\kappa,q)=\frac{f_{V}}{m_{V}^{2}}
\int\frac{d^{2}\ell<V_{i}^{\lambda_{i{}}}V_{f}^{\lambda_{f}}>}
{\ell^{2}}\exp(-\frac{|\ell-(\kappa+zq)|^{2}}{m_{V}^{2}z\overline{z}}),
\label{eq:general_impfact}
\end{equation}

The contractions of vertices for different helicities are:

\begin{equation}
<V_{i}^{1}V_{f}^{1}>=\ell^{*}(\ell-(\kappa+zq))(\frac{1}{{z^{2}}}+\frac{1}{\overline{z}^{2}})\label{eq:vertices}\end{equation}
\[
<V_{i}^{1}V_{f}^{0}>=\ell^{*}m_{V}(\frac{1}{z}-\frac{1}{\overline{z}})\]
\[
<V_{i}^{1}V_{f}^{-1}>=\ell^{*}(\ell-(\kappa-zq))^{*}\frac{2}{z\overline{z}}\]
In (\ref{eq:general_impfact}) the integration over $ \ell $ can be done
without further approximation, e.g. in the case
$\lambda_i=\lambda_f=1$ it leads to
$$ \int{d^2}\widetilde{\ell} \frac{(\widetilde{\ell}-\widetilde{\kappa})\widehat{\ell}{^*}}{\widetilde{\ell}{^2}} e^{-(\widetilde{\ell}-\widetilde{\kappa})^2}=\pi e^{-\widetilde{\kappa}^2},$$
$$\widetilde{\ell}=\frac{\ell}{m_{V}\sqrt{z\overline{z}}}, 
\widetilde{\kappa}=\frac{\kappa+zq}{m_{V}\sqrt{z\overline{z}}}$$
We obtain
\begin{equation}
\varphi^{1,1}(z,\kappa,q)=\pi f_{V}
z \overline{z}\exp(-\frac{|\kappa+zq|^{2}}{m_{V}^{2}z\overline{z}})
\left(\frac{1}{z^{2}}+\frac{1}{\overline{z}^{2}}\right)
\label{eq:integrated_impfact}\end{equation}
\[
\varphi^{1,0}(z,\kappa,q)=2 \pi f_{V}
m_{V}\frac{(\kappa+zq)^{*}}{|\kappa+zq|^{2}}(1-\exp(-\frac{|\kappa+zq|^{2}}{m_{V}^{2}z\overline{z}}))\]
\[
\varphi^{1,-1}(z,\kappa,q)=2 \pi f_{V}
m_{V}\frac{(\kappa+zq)^{*}{}{}{}^{2}}{|\kappa+zq|^{2}}
\left(\left(1+\frac{m_{V}^{2}z\overline{z}}
{|\kappa+zq|^{2}}\right)\exp(-\frac{|\kappa+zq|^{2}}
{m_{V}^{2}z\overline{z}})-\frac{m_{V}^{2}z\overline{z}}
{|\kappa+zq|^{2}}\right)\]

 Substituting into (\ref{eq:general_impfact}) we observe that
 there are terms contributing only in the vicinity of $|\kappa+zq|=0$
 or $|\kappa+\overline{z}q|=0$ corresponding to Landshoff-type
 kinematics. We have hard contributions, $\kappa<q$,  for
 $\lambda_{f}=0$ and $\lambda_{i}=-\lambda_{f},$ which for
 $z=O(1)$ can be written as
 \begin{equation}
 z\overline{z}\varphi_{4}^{1,0}|_{\kappa{\ll}q}=\pi f_{V}
 \frac{\kappa}{q^2}(2-\frac{1}{(z\overline{z})})
 \label{hardimpact} \end{equation}
 $$z\overline{z}\varphi_{4}^{1,-1}|_{\kappa{\ll}q}=2\pi f_{V} \frac{\kappa}{q^3}(-3+\frac{1}{(z\overline{z})})$$
 The singularities at the end points are spurious. Actually the
 integration over $z$ can be done with the result
 (\ref{eq:integrated_impfact}) without doing further
 approximations. However the result can be represented
 approximately by a z-integral with (\ref{hardimpact}) in the
 integrand and the range $\kappa/q < z <1-\kappa/q$.
  There is no hard chirally even contribution to the impact factor
 $\lambda_i=\lambda_f=1$ as the result of our particular choice
 of $\Psi ^{\gamma}$.
 There is only the Landshoff-type contribution, at
 $\kappa^{\prime}=\kappa+zq\ll q$
 \begin{equation}
 z\overline{z}\varphi_{4}^{1,1}|_{\kappa{\approx}q}=\pi f_{V} (z^2+\overline{z}^{2})
 {\exp}( {-\frac{|\kappa^{\prime}|^2}{m_{V}z\overline{z}}} )
 \label{softimpact} \end{equation}
 and the analogous one at $\kappa+\overline{z}q\ll q $.
 There are extra Landshoff-type contributions to the other
 helicities. In the case of ${\lambda}_f=0$ this results in a small
 contribution $O(q^{-5})$ to the amplitude and can be neglected.

\section{BFKL in double-logarithmic approximation.}
\setcounter{equation}{0}

Consider the BFKL equation \cite{BFKL} in the leading $\ln s$ approximation

\[
f(\omega,q,\kappa,\overline{\kappa})=f_{0}+\frac{g^{2}N}
{(2\pi)^{3}}\int\frac{d^{2}\kappa^{\prime}}
{|\kappa^{\prime}|^{2}|q-\kappa|^{2}}
K^{0}(\kappa^{\prime},\kappa,q)
f(\omega,q,\kappa^{\prime},\overline{\kappa})\]

\begin{equation} - \frac{g^{2}N} {(2\pi)^{3}}
{[\alpha(\kappa)+\alpha(q-\kappa)]f(\omega,q,\kappa,\overline{\kappa})},
\label{eq:evoltioneq} \end{equation}
with the bare kernel

\begin{equation}
K^{0}(\kappa^{\prime},\kappa,q)=
\frac{\kappa_{1}^{\prime}\kappa_{1}^{*}\kappa_{2}^{\prime *}\kappa_{2}+c.c}
{|\kappa_{1}-\kappa_{1}^{\prime}|^{2}}.
\label{eq:kernel}\end{equation}
We are going to simplify the equation in the double-log approximation, i.e.
we shall approximate the transverse momentum integrals in $\ln t $
approximation \cite{Fippel:yz}.
In double logarithmic approximation the gluon trajectory function
can be written as
\[
\alpha(\kappa)\approx\frac{g^{2}N}{(2\pi)^{3}}\int_{\mu^{2}}^{|\kappa|^{2}}
\frac{d^{2}\kappa^{\prime}}{|\kappa^{\prime}|^{2}}=
N(\frac{g^{2}}{4\pi})\frac{1}{2\pi}\ln\frac{|\kappa|^{2}}{\mu^{2}}\]
Taking also into account running of the coupling we get
\begin{equation}
\alpha(\kappa)\approx
N\int_{\mu^{2}}^{|\kappa|^{2}}\frac{d|\kappa^{\prime}|^{2}}
{|\kappa^{\prime}|^{2}}\frac{\alpha_{s}(|\kappa^{\prime}|^{2})}{2\pi}\equiv
N\xi(\kappa)\label{eq:alpha}\end{equation} According to  Section 2 we
want to consider two different kinematical cases:

1. $\kappa_{2}^{\prime}\approx \kappa_{2}\approx q\gg \kappa_{1}^{\prime}\gg
\kappa_{1}$

The evolution kernel in this region can be approximated as:

\begin{equation}
K^{0}(\kappa^{\prime},\kappa,q)=
\frac{\kappa_{1}^{*}}{\kappa_{1}^{\prime*}}|q|^{2}+c.c
\label{eq:hard_kernel}\end{equation}
Replacing $f=\kappa\widetilde{f}$ the  equation becomes
\[
\widetilde{f}=\widetilde{f}_{0}+\frac{g^{2}N}
{(2\pi)^{3}\omega}\int_{\kappa^{2}}^{|\overline{\kappa}|^{2}}
\frac{d^{2}\kappa^{\prime}}{|\kappa^{\prime}|^{2}}\widetilde{f}
(\omega,\kappa^{\prime};\overline{\kappa})-...\]
\begin{equation}
=\widetilde{f}_{0}+
N\int_{\xi(\kappa)}^{\xi(\overline{\kappa})}
d\xi^{\prime}\widetilde{f}(\xi^{\prime},\overline{\xi})-...
\label{eq:hard_evolution}\end{equation}

2. $\kappa_{1}\approx \kappa_{1}^{\prime}$$\approx \kappa_{2}\approx
\kappa_{2}^{\prime}$$\approx q$$\gg|\kappa_{1}-\kappa_{1}^{\prime}|.$ 
We parametrize 
 \[
\kappa_{1}=zq+\widetilde{\kappa};\kappa_{2}=\overline{z}q-\widetilde{\kappa}\]
\[
\kappa_{1}^{\prime}=zq+\widetilde{\kappa}^{\prime};\kappa_{2}^{\prime}=
\overline{z}q-\widetilde{\kappa}^{\prime}\]
\[
q\gg\widetilde{\kappa}^{\prime}\gg\widetilde{\kappa}\]
The kernel in this kinematics is approximately 
$K^{0}= \frac{2|q|^{4}}{|\widetilde{\kappa}^{\prime}|^{2}}$
Then the  equation can be written as
\[
f=f_{0}+\frac{2g^{2}N}{(2\pi)^{3}\omega}
\int_{|\mu|^{2}}^{|\overline{\kappa}|^{2}}
\frac{d^{2}\widetilde{\kappa}^{\prime}}{\widetilde{|\kappa}^{\prime}|^{2}}
f(\widetilde{\kappa}^{\prime},\overline{\kappa})-...\]
\begin{equation}
=f_{0}+\frac{2N}{\omega}\int_{\xi(\kappa)}^{\xi(\overline{\kappa})}
d\xi^{\prime}f(\xi^{\prime},\overline{\xi})
\label{eq:soft_evolution}\end{equation}

In our approximation we can solve the evolution equations 
for both cases.

In the first case, substituting
$\widetilde{f}\rightarrow\frac{\widetilde{f}}{\omega}$, we rewrite
the equation (\ref{eq:hard_evolution}) as
\[
\left(\omega+N\xi(\kappa)+N\xi(q)\right)
\widetilde{f}(\omega,q,\xi,\overline{\xi})=
\widetilde{f}_{0} +
N\int_{\xi(\kappa)}^{\xi(\overline{\kappa})}d\xi^{\prime}
\widetilde{f}(\omega,\xi^{\prime},\overline{\xi})\]

and by another substitution
$ \hat {f} (\omega,q,\xi,\overline{\xi})
\equiv\left(\omega+N\xi(\kappa)+N\xi(q)\right)
\widetilde{f}(\omega,q,\xi,\overline{\xi}) $
we transform the equation into

\begin{equation}
\hat {f}(\omega,q,\xi,\overline{\xi})=
\widetilde{f}_{0} +N\int_{\xi(\kappa)}^{\xi(\overline{\kappa})}d\xi^{\prime}
\frac{\hat {f} (\omega,q,\xi^{\prime},\overline{\xi})}
{\omega+N\xi^{\prime}(\kappa)+N\xi(q)}
\label{eq:hard_integral}\end{equation}
The solution is
\begin{equation}
\widetilde{f}(\omega,q,\kappa,\overline{\kappa})=
\widetilde{f}_{0}
\left\{
\frac{1}{\omega+N\xi(\kappa)+N\xi(q)}+\frac{N\xi(\overline{\kappa})-N\xi(\kappa)}{\left(\omega+N\xi(\kappa)+N\xi(q)\right)^{2}}\right\}
\label{eq:hard_solution}\end{equation}

Carrying out the Mellin transformation of this expression $${\cal
G}_{h}(s,q,\kappa,\overline{\kappa})=
\int_{-i\infty}^{i\infty}\frac{d\omega}{2\pi
i}\widetilde{f}(\omega,q,\kappa,\overline{\kappa})\left(
\frac{s}{|q|^{2}}\right)^{\omega}$$
we obtain the gluon exchange Green function of the scattering amplitude in
double-logarithmic approximation.
\begin{equation}
{\cal {\cal G}}_{h}
(s,q,\kappa,\overline{\kappa})=\left(\frac{s}{|q|^{2}}
\right)^{-N(\xi(\overline{\kappa})+N\xi(\kappa))}
(1+N\ln\frac{s}{q^{2}}(\xi(\overline{\kappa})-
\xi(\kappa))
\label{eq:hard_amplitude}\end{equation}


In the other case of interest the evolution equation
(\ref{eq:soft_evolution}) is written as

\begin{equation}
f(\omega,q,\xi,\overline{\xi})=
\frac{f_{0}}{\omega}+\frac{2N}{\omega}
\int_{\xi(\kappa)}^{\xi(\overline{\kappa})}
d\xi^{\prime}f(\xi^{\prime},\xi(\overline{\kappa}))-N(\xi(zq)+
\xi(\overline{z}q))f(\xi,\xi(\overline{\kappa}))
\label{eq:soft_integral}\end{equation}

Proceeding analogously to the previous case we get the
differential equation
\begin{equation}
\frac{d}{d\xi}f(\omega,q,\xi,\overline{\xi})=
-\frac{2Nf(\omega,q,\xi,\overline{\xi})}
{\omega+N\xi(zq)+N\xi(\overline{z}q)}
\label{eq:soft_differential}\end{equation}

The solution of this equation is
\begin{equation}
f(\omega,q,\xi(\kappa),\xi(\overline{\kappa})=\frac{f_{0}}
{(\omega+N\xi(zq )+N\xi(\bar z q))}
\exp\left(\frac{2N(\xi(\overline{\kappa})-\xi(\kappa))}
{\omega+N\xi(zq)+N\xi(\bar{z}q)}\right)
\label{eq:soft_solution}\end{equation}

Carrying out the Mellin transformation we obtain the gluon
exchange Green function
\begin{equation}
{\cal {\cal G}}_{L}
(s,q,\kappa,\overline{\kappa})=\left(\frac{s}{|q|^{2}}
\right)^{N(2\xi(\overline{\kappa})-2\xi(\kappa)-\xi(zq)-\xi(\overline{z}q))}
\label{eq:soft_amplitude}
\end{equation}

\section{Helicity amplitudes }
\setcounter{equation}{0}

At large t the diffractive exchange interacts with a single quark in the disintegrating proton.
We write down the helicity amplitudes of diffractive scattering on a quark. In this case the proton impact factor reduces to a constant and the coupling of the exchange to the disintegrating proton
does not influence the $t$-dependence.
We obtain the diffractive $\gamma V$ amplitudes in terms of a sum
of the hard and Landshoff-type contributions, each of them has the
form
\begin{equation}
M^{{\lambda}_i{\lambda}_f}=is{\int}{d^2}\kappa dz
z\overline{z}{\varphi}_{4}^{{\lambda}_i{\lambda}_f}(\kappa,q,z)\frac{{\cal
{\cal G}}(s,q,\kappa,\overline{\kappa})}{|\kappa|^2|\kappa-q|^2} \label{eq:main}
\end{equation}
For comparison we consider first the amplitudes with simple
two-gluon exchange, i.e. we substitute ${\cal {\cal G}}=1$, and denote this by an additional subscript 1. The
hard contribution to the cases ${\lambda}_f=0$, and
${\lambda}_i=-{\lambda}_f$ can be calculated in this case without
further approximation with the results
\begin{equation}
M^{1,0}_{1,h}=-2C m_{V}\pi^2
(4-\frac{\pi^2}{3})\frac{q}{t^{2}}\label{eq:hard_ampl}\end{equation}
\[
M^{1,-1}_{1,h}=C \frac{2\pi^2}{t^2} (\frac{2\pi^2}{3}-8). \]
The factor C denotes $ C=is\frac{2}{3} \alpha_{s}^{2}eQ_{q}f_{V}\Phi_{P}$. The results for the hard contributions essentially coincide with the ones obtained in \cite{Ivanov:2000uq} for the corresponding chiral even contributions.

The Landshoff-type contributions for 2-gluon case are
\[ M^{1,1}_{1,L}=\int
d^{2}\kappa^{\prime}\int_{0}^{1}dzz\overline{z}
\varphi_{4}^{11}(z,\kappa,q)\frac{1}{|\kappa|^{2}|q-\kappa|^{2}}\]
\[
\approx C \int
d^{2}\kappa^{\prime}\int_{0}^{1}dz\frac{\pi}{|q|^{4}}\exp(-\frac{|\kappa^{\prime}|^{2}}{m_{V}^{2}z\overline{z}})\left(\frac{1}{z^{2}}+\frac{1}{\overline{z}^{2}}\right)\]
\begin{equation}
=C \frac{2 \pi^{2}}{|q|^{4}}\int_{m_{V}/q}^{1-m_{V}/q}
(\frac{1}{z\overline{z}}-2)dz.
\label{eq:int_transv_amplitude}\end{equation}


There is also contribution to double spin-flip amplitude which has
to be added to the hard contribution.
\begin{equation}
M^{1,-1}_{1,L}=C \frac{2\pi^{2}}{|q|^{4}}
\int_{m_{V}/q}^{1-m_{V}/q}(\frac{1}{z\overline{z}}-3)dz
\end{equation}
$M^{1,0}_{1,L}$ is small compared to the hard contribution.

 Now we evaluate the amplitudes including the double-log
 approximation to the BFKL evolution. We substitute
${\cal G}(s,k,q)$ by ${\cal {\cal G}}_{h}$ (\ref{eq:hard_amplitude}) for the
hard contribution from the region $|\kappa| \ll |q|,|\kappa-q| \ll |q|$  and by ${ \cal {\cal G}}_{L}($ \ref{eq:soft_amplitude})
 for the Landshoff-type  contributions from the regions $|\kappa+zq| \ll |q|$,$|\kappa+\overline{z}q| \ll |q|$. The upper transverse momentum $\overline{\kappa}$ is of order $q$ in ${\cal {\cal G}}_{h}$
 and of order $ min(z,\overline{z}) q$ in ${\cal {\cal G}}_{L}$. The integration is dominated by
 $k \approx m_{V}$. Therefore, the double-logarithmic interaction in the gluon exchange tends to suppress the hard contribution. There is no suppression in the Landshoff-type contribution for $z=O(1)$, but the end-point contribution are damped.

This effects the t dependence of the amplitudes over a large t range. The combined effect of the end-points and ${\cal {\cal G}}_{L}$ leads to a clear flattering of the t dependence compared to the naive $1/|t|^2$ in the non-flip amplitude. In the small range $3<|t|<10 {GeV}^2$ the modification in the other amplitudes is small.

\section{Numerical evaluation and discussion.}
\setcounter{equation}{0}

The helicity amplitudes calculated above allow to evaluate the
$t$ dependence of the vector meson production cross section and
of the angular-decay coefficients \cite{Schilling:1973ag}. For the latter we
use the relations,
$$
r^{04}_{00}\propto \frac{1}{N}|M^{10}|^{2}
$$
\begin{equation}
r_{10}^{04}\propto \frac{1}{2N}(M^{10 *} M^{11}+M^{10 *}M^{1-1})
\end{equation}
$$
r_{1-1}^{04}\propto \frac{1}{N}(M^{1-1*}M^{11})
$$
$$
N=|M^{10}|^{2}+|M^{11}|^{2}+|M^{1-1}|^{2}
$$
Our estimates are done for large $t$, therefore it makes sense to extract
results for values of $|t|$ above 3 $GeV^2$.

With our estimates we do not predict the normalization of the cross section. Besides of this,
no parameters are fitted.

The predicted $t$-dependence of the cross section agrees reasonably with the data \cite{Chekanov:2002rm}. In Fig.2 we show how the double-log interaction in the exchange improves the t-dependence.

We compare also with the $t$-dependence resulting from the amplitudes given in \cite{Ivanov:2000uq};
it deviates from the experimental behaviour. We conclude that the combined effect of large-size dipole contributions and of exchange interaction improves the naive power behaviour in $t$ in  agreement with experiment.

 A more detailed description of the t dependence has been achieved in \cite{Poludniowski:2003yk},
 by including the leading ln s BFKL solution for the diffractive exchange and fitting some parameters. However, the angular decay coefficients obtained there are not in full agreement with
 the data, in particular $r_{10}^{04}$ turns out to have the opposite sign. It has been pointed out that  the right sign could be achieved by increasing the chirally-odd contribution parametrized
 there by the consituent quark mass, giving the latter an unreasonable large value.

 In Fig.3 we show our results on the angular-decay coefficients in comparison with the data and also with the predictions calculated from the results of \cite{Ivanov:2000uq}. It is clear that these coefficients are sensitive to the detailed structure of the amplitudes which is not resolved
 in the cross section.

We have seen that the relative magnitude of the amplitudes arising from helicity dependence of the impact factors, is influenced essentually by end-point contributions and also by the interaction of the exchanged gluons. Neglecting these effects would result in the dominance  of the longitudinal polarization of the produced vector meson. We obtain that in the experimetally accesible t-range non-flip and single-flip are of approximately the same magnitude and $t$-dependence. The double-flip amplitude is about an order of magnitude smaller, due to partial cancellation of soft and hard contributions. The data on $r_{00}^{04}$ suggests that the non-flip amplitude is even larger, exceeding the single-flip amplitude by a factor 3 to 4.

Besides of this our results on the angular decay coefficient are in qualitative agreement with experiment and also with the results derived from \cite{Ivanov:2000uq}, in particular $r_{10}^{04}$
agrees in sign with experiment. In the latter paper the inclusion of a sizable chirally odd contribution was essential for the qualitative agreement with experiment, in particular for the observed  dominance of the transverse polarization of the produced vector meson.

Our amplitudes do not include a chirally odd contribution. In this way we have shown that the end-point and exchange interaction account partially for the correction achieved in \cite{Ivanov:2000uq} by the addition of a chirally odd component.

In this paper we have emphasized the role of contributions with relatively
large dipole size. We have estimated in the double log approximation the
dominant effect of the exchanged gluon interaction. Without including a chirally odd contribution we obtaine that the transverse
and longitudinal production rates are of the same size in the $t$ range of
interest. The angular decay distribution data suggest a stronger
enhancement of the transverse production, leaving room for a sizable
chirally odd contribition to the photon and meson wave functions.

A lesson to be drawn from this study is that the large $t$ transition
impact factor has, besides of the small-dipole part, which can be
parametrized by photon and meson distribituion amplitudes, a part with
relatively large dipole sizes. For the latter one has to introduce 
complete light-cone wave functions of involved particles; their values in the
vicinity of vanishing dipole size are not sufficient. The calculation of
diffractive large $t$ amplitudes requires the corresponding additional
non-perturbative input. 

\newpage
\begin{figure}[htbp]
\begin{center}
\vspace{-0.cm}
\epsfig{file=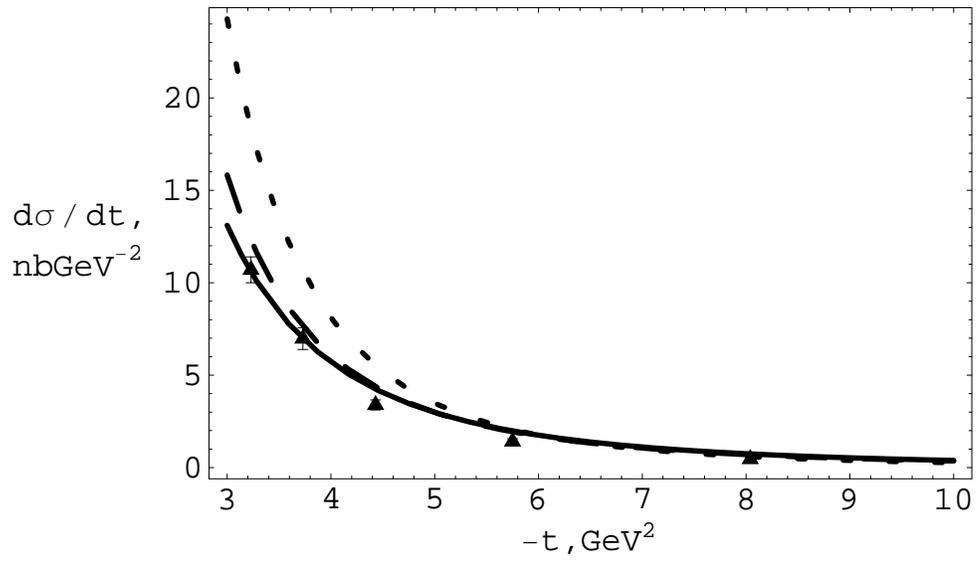,width=12.9cm}
\end{center}
\caption{The diffractive $\gamma V$  cross-section. 2-gluon picture without double-log corrections - dense curve, with included double-logs - dashed curve, the result of \cite{Ivanov:2000uq} - dotted curve. Experimental points - ZEUS 2002.}
\label{fig:cross}\end{figure}
\begin{figure}[htbp]
\begin{center}
\vspace{-0.cm}
\epsfig{file=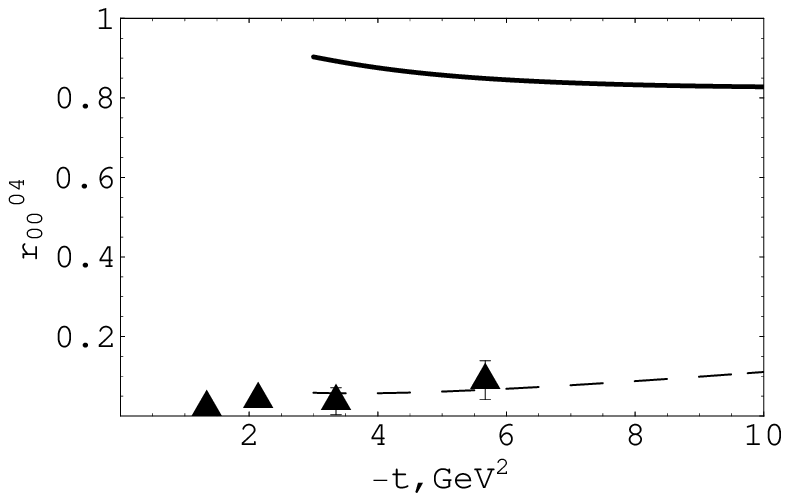,width=9.7cm}
\epsfig{file=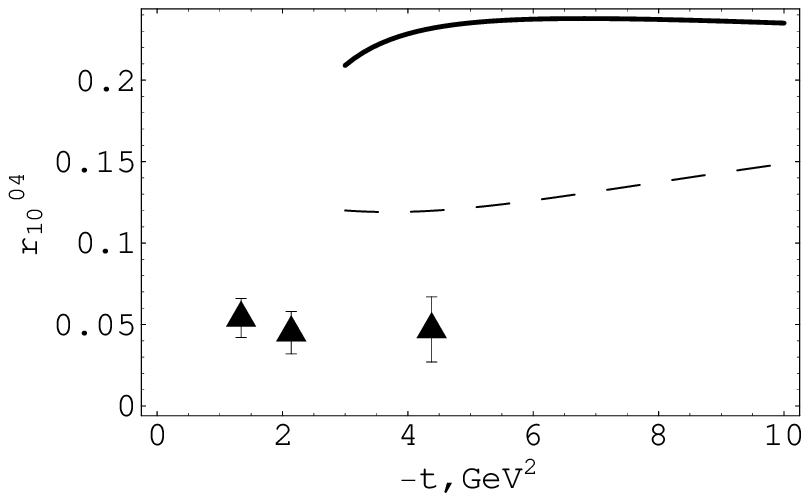,width=9.7cm}
\epsfig{file=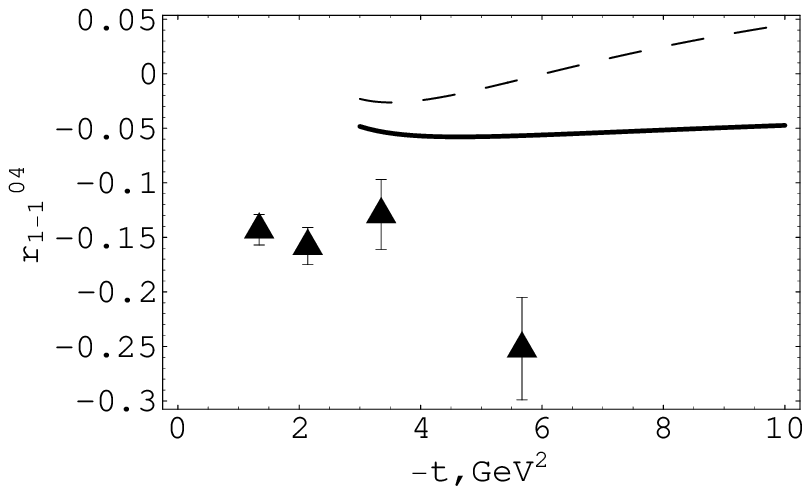,width=9.7cm}
\end{center}
\caption{
The angular-decay coefficients $r_{00}^{04}$, $r_{10}^{04}$, $r_{1-1}^{04}$. The dotted curve - results of \cite{Ivanov:2000uq}, the dense curve - present calculation.Experimental points - ZEUS 2002.} \label{fig:r} \end{figure}


\end{document}